\documentclass[11pt]{article}
\usepackage[a4paper, total={6in, 8in}]{geometry}
\usepackage{epsfig,enumerate,verbatim}
\usepackage{amsmath}
\usepackage{amsfonts}
\usepackage{amssymb}
\usepackage{amstext}
\usepackage{amsthm}
\usepackage{dsfont}
\usepackage{mathrsfs}
\usepackage{cite}

\begin{document}        
\title{\bf{Non-Relativistic Anti-Snyder Model and Some Applications}}
\author{C. L. Ching\footnote{Email: phyccl@nus.edu.sg (corresponding author)}, C. X. Yeo\footnote{Email: kencx89@hotmail.com} and W. K. Ng\footnote{Email: phynwk@nus.edu.sg}}

\date{15$^{\text{th}}$ Jan 2016}

\maketitle

\begin{center}
{Department of Physics,\\}
{National University of Singapore,\\}
{Kent Ridge, Singapore.}
\end{center}

\begin{abstract}
\noindent We examine the (2+1)-dimensional Dirac equation in a homogenous magnetic field under the non-relativistic anti-Snyder model which is relevant to doubly/deformed special relativity (DSR) since it exhibits an intrinsic upper bound of the momentum of free particles. After setting up the formalism, exact eigen-solutions are derived in momentum space representation and they are expressed in terms of finite orthogonal Romanovski polynomials. There is a finite maximum number of allowable bound states $n_{\text{max}}$ due to the orthogonality of the polynomials and the maximum energy is truncated at $n_{\text{max}}$. Similar to the minimal length case, the degeneracy of the Dirac-Landau levels in anti-Snyder model are modified and there are states that do not exist in the ordinary quantum mechanics limit $\beta\rightarrow 0$. By taking $m\rightarrow 0$, we explore the motion of effective massless charged fermions in graphene like material and obtained a maximum bound of deformed parameter $\beta_{\text{max}}$. Furthermore, we consider the modified energy dispersion relations and it's application in describing the behavior of neutrinos oscillation under modified commutation relations.       
\end{abstract}

{\bf Keywords}: Modified Commutation Relations; Snyder and anti-Snyder Algebra; Dirac-Landau Problem; Modified Energy Dispersion Relations; Modified Neutrinos Oscillation; Quantum Gravity Phenomenology. 

\section{Introduction}

In 1947, H. Snyder \cite{snyder} proposed a modification to the canonical commutation relations of position and momentum operators in the form of $[\widehat{X}_\mu, \widehat{X}_\nu] \neq 0$. Spacetime became Lorentz-covariantly non-commutative, but the modification of commutation relations increased the ordinary Heisenberg's uncertainty, such that a smallest possible resolution of spacetime structures was introduced. It was the first to show that a fundamental minimal length scale needs not be in conflict with Lorentz invariance, i.e. Snyder algebra is invariant under the Lorentz group \cite{snyder}. 

The existence of absolute minimal length is predicted by various approaches to unify quantum mechanics with gravity in the ultraviolet sector. For example, perturbative string theory \cite{amati1, gross}, quantum gravity \cite{garay} and black holes physics \cite{maggiore} suggest a generalized uncertainty principle \cite{kempf, witten} which implies a minimum uncertainty in position \cite{maggiore, amati2}. For review see \cite{sabine, sprenger1} and experimental relevance in \cite{optics, bar, oscillator}. 

The relativistic modified quantum algebra introduced by Snyder is based on modified commutation relation (MCR) \cite{snyder}
\begin{eqnarray}
\left[\widehat{X}_\mu, \widehat{X}_\nu \right] &=& i \hbar\ \tilde{\beta} \widehat{J}_{\mu\nu}\ ; \hspace{0.25cm}[\widehat{P}_\mu, \widehat{P}_\nu] = 0 \nonumber\\
\left[\widehat{X}_\mu, \widehat{P}_\nu \right] &=& i \hbar \bigl(\eta_{\mu\nu} + \tilde{\beta} \widehat{P}_{\mu} \widehat{P}_{\nu}\bigr)\label{MCR1}
\end{eqnarray}
where $\tilde{\beta}$ is a coupling constant usually assumed to be of the scale of the square of the Planck length and dimensionally $[\tilde{\beta}]=[\text{momentum}^{-2}]$. $J_{\mu\nu}$ are the generators of the Lorentz symmetry and $(\mu,\nu= 0,1,2,3)$ are the spacetime indices. The physical predictions of MCR\eqref{MCR1} depend strongly on the signature of the coupling constant $\tilde{\beta}$. In literature, $\tilde{\beta}>0$ is known as Snyder model while $\tilde{\beta}<0$ is called anti-Snyder model \cite{mig1}. Interestingly, Snyder model can be viewed as generalization of the uncertainty relations that imply a minimum bound for the uncertainty in position \cite{kempf,das,spectra} and the discreteness of the spectra of area and volume\cite{romero} similar to that proposed by loop quantum gravity \cite{lqg1, lqg2}. On the other hand, anti-Snyder model ($\tilde{\beta}<0$) exhibits an upper bound on the mass of free particles \cite{mig2} and thus can be considered as particular realization of DSR axioms \cite{dsr}. Also, Snyder/anti-Snyder model is highly related to de-Sitter/anti de-Sitter spacetime for momentum space \cite{snyder}. For details of applications and interpretation of Snyder/anti-Snyder model, see \cite{mig1, snyderapp} and review \cite{livine}.

Our manuscript is organized as follows: In Sect.(2), we provide the formalism of the deformed quantum mechanics under anti-Snyder model in momentum representation and discuss the self-adjointness condition of the physical operators. In Sect.(3), following \cite{menculini-landau} we study the Dirac-Landau problem under anti-Snyder model with maximum momentum cutoff. Both eigenspectrum and eigenfunction are exactly determined in terms of finite orthogonal Romanovski polynomials. We also consider the potential application of the model to massless charged fermion, e.g. electronic Landau levels in graphene like 2-dimensional systems. In Sect.(4), we consider the modified energy dispersion relation (MDR) and apply it to neutrinos oscillation. By removing the universality of the deformed parameter (flavour dependence), it is possible to have neutrino oscillation for massless or degenerate mass neutrinos. Finally, we summarize our main results in the concluding section and the appendix contains a brief discussion of the Romanoski polynomials and their properties.   

\section{Quantum Mechanics with Anti-Snyder Model in Momentum Representation}

The non-relativistic version of anti-Snyder model \eqref{MCR1} can be realized  as a deformed quantum mechanics \cite{mig1} by a set of 3-dimensional MCR given as following:
\begin{eqnarray}
\left[\widehat{X}_{i}, \widehat{X}_{j}\right] &=&-i \hbar\ \beta (1-\beta \widehat{P}^2) \widehat{L}_{ij} = -i \hbar\ \beta \bigl(\widehat{X}_{i} \widehat{P}_{j}- \widehat{X}_{j} \widehat{P}_{i}\bigr)\nonumber\\
\left[\widehat{X}_{i}, \widehat{P}_{j}\right] &=& i \hbar\ \bigl(\delta_{ij} - \beta \widehat{P}_{i} \widehat{P}_{j}\bigr)\ ; \hspace{0.25cm}\left[\widehat{P}_{i}, \widehat{P}_{j}\right] = 0 \label{MCR2}\\
\left[\widehat{X}_{i}, \widehat{L}_{j}\right] &=& i \hbar \epsilon_{ijk} \widehat{X}_{k}\ ; \hspace{0.25cm}\left[\widehat{P}_{i}, \widehat{L}_{j}\right] = i \hbar \epsilon_{ijk} \widehat{P}_{k}\ ; \hspace{0.25cm}\left[\widehat{L}_{i}, \widehat{L}_{j}\right] = i \hbar \epsilon_{ijk} \widehat{L}_{k}\nonumber
\end{eqnarray}
where Latin indices $(i,j = 1,2,3)$ are the spatial indices, $\tilde{\beta} = - \beta < 0$ is the deformed parameter with dimension of inverse square of momentum $\left[\beta\right]=\left[\text{momentum}^{-2}\right]$ and typically assumed to be universal and positive definite $\beta > 0$. Angular momentum vector is defined as $\widehat{L}_{i}:= \frac{1}{1-\beta \hat{p}^2}\ \epsilon_{ijk} \widehat{X}_{j} \widehat{P}_{k}$ and $\widehat{L}_{ij}=\epsilon_{ijk} \widehat{L}_{k}$. Anti-Snyder algebra above satisfies Jacobi identity exactly and the spatial degree of freedom we considered is a non-commutative geometry. In contrast to the string theory inspired MCR \cite{kempf, laynam} (Snyder's algebra and Kempf's algebra) which exhibits a minimal length, anti-Snyder model contains an upper bound in modulus of momentum that would give a realization of the DSR axioms \cite{mig1, dsr}. In the literatures, there are other form of MCR's proposed to study the idea of DSR, for example see \cite{das, das-rel, jiz, cutoff, pedrammax, spectra, potwell, coherent, rwe}. From \eqref{MCR2}, the generalized uncertainty principle (GUP) is
\begin{eqnarray}
(\Delta \widehat{X}_i) (\Delta \widehat{P}_j) &\geq& \frac{1}{2}\bigl|\langle \bigl[\widehat{X}_i, \widehat{P}_j\bigr]\rangle\bigr| = \frac{\hbar}{2} \delta_{ij}\left(1 - \beta \sum_{k=1}^3\Bigl((\Delta \widehat{P}_k)^2 - \langle\widehat{P}_k\rangle^2\Bigr)\right)
\end{eqnarray}
where it can be rewritten in one-dimension as 
\begin{eqnarray}
\frac{2(\Delta \widehat{X})}{\hbar} &\geq& \frac{1 -\beta \langle \widehat{P}\rangle^2}{(\Delta \widehat{P})}- \beta (\Delta \widehat{P}).
\end{eqnarray}
We see that there is no minimal position uncertainty\footnote{The minimal position uncertainty is usually interpreted as the minimum length scale induced by the gravitational effects from UV sector on the various quantum mechanical systems.} but there is a finite momentum uncertainty \cite{mig1}, e.g. $(\Delta \widehat{P})= \frac{1}{\sqrt{\beta}}$ when $(\Delta \widehat {X})=0$. 

Furthermore we can choose to represent the position and momentum operators in momentum space by $\widehat{P}_{i} = \hat{p}_{i}$, \ $\widehat{X}_{i} = (1 - \beta \hat{p}^{\ 2}) \hat{x}_{i}$\  where $\hat{x}_{i},\ \hat{p}_{i}$ are the low energy canonical variables satisfying the usual Heisenberg algebra
\begin{eqnarray}
\left[\hat{x}_{i}, \hat{p}_{j}\right] &=& i \hbar \delta_{ij} \label{canon}
\end{eqnarray}
with $\hat{x}_{i} : = i \hbar \frac{\partial}{\partial p_{i}}$. Thus, the representation of the anti-Snyder algebra \eqref{MCR2} is exactly given by 
\begin{eqnarray}
\widehat{P}_i\ \psi(P) &=& p_i\ \psi(p)\label{rep.P}\\
\widehat{X}_i\ \psi(P) &=& i \hbar f(P) \frac{\partial \psi(P)}{\partial P_i} = i \hbar \bigl(1 - \beta p^2\bigr) \frac{\partial \psi(p)}{\partial p_i} \label{rep.X}
\end{eqnarray}
where we have defined parity invariant deforming factor $f(-P) = f(P)$. Both operators \eqref{rep.P} and \eqref{rep.X} are defined to act on the dense domain of $C^{\infty}([-\beta^{-1/2}, \beta^{-1/2}])$. The deformed measure and completeness relation are given by
\begin{eqnarray}
1_f &\equiv& \int_{-\frac{1}{\sqrt{\beta}}}^{\frac{1}{\sqrt{\beta}}}\ \frac{d^3 \vec{p}}{f}\ |\vec{p}\rangle\langle \vec{p}| = \int_{-\frac{1}{\sqrt{\beta}}}^{\frac{1}{\sqrt{\beta}}}\frac{d^3 \vec{p}}{1- \beta p^2}\ |\vec{p}\rangle\langle \vec{p}| ,\label{measure}
\end{eqnarray}
and the inner product of momentum eigenstate is
\begin{eqnarray}
\langle \vec{p}|\vec{p}\ '\rangle &=& \bigl(1- \beta \vec{p}^2\bigr)\ \delta^3(\vec{p}-\vec{p}\ ').\label{inner1}\\
\Rightarrow\ \langle \phi|\psi\rangle_f &\equiv& \int_{-\frac{1}{\sqrt{\beta}}}^{\frac{1}{\sqrt{\beta}}} \frac{d^3 \vec{p}}{f} \phi^*(\vec{p})\psi(\vec{p})  = \int_{-\frac{1}{\sqrt{\beta}}}^{\frac{1}{\sqrt{\beta}}} \frac{d^3 \vec{p}} {1- \beta p^2} \phi^*(\vec{p})\psi(\vec{p})\, .\label{inner2}
\end{eqnarray}
The choice of the deformed measure in \eqref{inner2} is crucial as it has to ensure the symmetricity of the position operator that can be seen from \eqref{herX} in the next section. We require the intrinsic momentum cutoff $ -1/\sqrt{\beta}< \vec{p} < 1/\sqrt{\beta}$ to be implemented in order to make the measure positive definite. The convergence of the integral in \eqref{inner2} allows us to single out the physical Hilbert space 
\begin{eqnarray}
\mathcal{H} &=& \mathcal{L}^{2}\Bigl[\Bigl(\frac{-1}{\sqrt{\beta}}, \frac{1}{\sqrt{\beta}}\Bigr), \frac{d^3 p}{f(p)}\Bigr]
\end{eqnarray}
i.e. the space of wavefunctions $\psi(\vec{p})$ such that
\begin{eqnarray}
\langle \psi|\psi\rangle_f &\equiv& \int_{-\frac{1}{\sqrt{\beta}}}^{\frac{1}{\sqrt{\beta}}} \frac{d^3 \vec{p}}{1- \beta p^2} |\psi(\vec{p})|^2 < \infty.
\end{eqnarray}
These directly imply that as $\vec{p}\to \pm 1/{\sqrt{\beta}}$, we have to impose the fall-off condition of wave function $\psi(\vec{p}) \to 0$. 

\subsection{Symmetricity and Self-Adjointness Conditions of Physical Operators}

We require $\widehat{X}_i$ and $\widehat{P}_i$ to be symmetric in the dense domain of $C^{\infty}([-\beta^{-1/2}, \beta^{-1/2}])$. The operator $\widehat{P}_i=\hat{p}_i$ is manifestly Hermitian in \eqref{rep.P}. The symmetry of $\widehat{X}_i$ can be deduced from,
\begin{eqnarray}
\left(\langle \phi|\widehat{X}_i|\psi \rangle \right)_f^{*}
&=& \left(\int_{\frac{-1}{\sqrt{\beta}}}^{\frac{1}{\sqrt{\beta}}} \frac{d^3 \vec{p}}{1- \beta p^2}\ \phi^{*}(\vec{p}) \left[i\hbar\Bigl(1-\beta
p^2\Bigr) \frac{\partial}{\partial p_i} \right] \psi(\vec{p}) \right)^{*}\nonumber\\
&=& \int_{\frac{-1}{\sqrt{\beta}}}^{\frac{1}{\sqrt{\beta}}} \frac{d^3 \vec{p}}{1-\beta p^2}\psi^{*}(\vec{p}) \left[i\hbar \Bigl(1-\beta p^2\Bigr) \frac{\partial}{\partial p_i} \phi(\vec{p}) \right] + B(\vec{p}),\nonumber\\
&=&\langle \psi|\widehat{X}_i|\phi\rangle_f + B(\vec{p}) \label{herX}
\end{eqnarray}
where we have performed integration by parts and obtained the boundary term 
\begin{eqnarray}
B(\vec{p}) &=& -i\hbar \Bigl(\psi^{*}(\vec{p}) \phi(\vec{p}) \Bigr) \Bigr|_{\frac{-1}{\sqrt{\beta}}}^{\frac{1}{\sqrt{\beta}}}. \label{bc1}
\end{eqnarray} 
Using the measure $d^3 \vec{p}/(1- \beta p^2)$ in \eqref{herX}, the position operator is symmetric if and only if the boundary terms $B(\vec{p})$ vanish. Although we ensure the boundary terms \eqref{bc1} to vanish by the appropriate fall-off condition of momentum wavefunction, i.e. to require $\phi \bigl(\vec{p}\to \pm\frac{1}{\sqrt{\beta}}\bigr) = 0$, there is no constraint on the conjugate dual $\psi^{*}(\vec{p})$. In general $\psi^{*}(\vec{p})$ can be arbitrary. This implies that although the position operator $\widehat{X}_i$ and its Hermitian conjugate $\widehat{X}^{\dagger}_i$ have the same formal expression, their domain may not be equivalent, i.e 
\begin{eqnarray}
\textit{D}(\widehat{X}_i) &=& \left\{\phi\in\mathcal{L}^{2} \Bigl(\frac{-1}{\sqrt{\beta}}, \frac{1}{\sqrt{\beta}}\Bigr); \phi\Bigl(\frac{-1}{\sqrt{\beta}}\Bigr) = \phi\Bigl(\frac{1}{\sqrt{\beta}}\Bigr) = 0\right\}\\ \label{dom1}
\textit{D}(\widehat{X}^{\dagger}_i) &=& \left\{\psi\in\mathcal{L}^{2}\Bigl(\frac{-1}{\sqrt{\beta}}, \frac{1}{\sqrt{\beta}}\Bigr);\text{no restriction on $\psi$}\right\}.\label{dom2}
\end{eqnarray}

To study the self-adjointness property of the position operator, one can apply the von Neumann's theorem \cite{Akh}. First, we solve the eigenvalue problem for the position operator with eigenvalues $\pm i\lambda$, where $\lambda>0$ is a real constant,
\begin{eqnarray}
\widehat{X}\phi_{\pm}(p)=i\hbar \bigl(1-\beta p^2\bigr)\frac{\partial}{\partial p}\phi_{\pm}(p)=\pm i\lambda \phi_{\pm}(p).\label{pos1}
\end{eqnarray}
A simple integration gives 
\begin{eqnarray}
\phi_{\pm}(p) = \phi_{\pm}(0) \exp\Bigl(\frac{\pm \lambda \tanh^{-1}(\sqrt{\beta} p)}{\hbar\sqrt{\beta}}\Bigr).
\end{eqnarray}
We see that $\phi_{\pm}(p)$ is not normalizable within the domain $p\in\Bigl(\frac{-1}{\sqrt{\beta}}, \frac{1}{\sqrt{\beta}}\Bigr)$ for both $\pm i\lambda$ cases. Hence the deficiency indices\footnote{The number of the possible self-adjoint extension is given by $n_{+}*n_{-}$, where $n_{+}(n_{-})$ are the deficiency indices \cite{Akh}.} \cite{Akh, Araujo} of position operator $\widehat{X}$ are $(0, 0)$. As a consequence, the position operator is indeed self-adjoint beside merely symmetric by Von Neumann's theorem.

By the same token, we have for momentum operator in position space
\begin{eqnarray}
\widehat{P}\phi_{\pm}(x)=-i\hbar\frac{\partial}{\partial x}\phi_{\pm}(x)=\pm i\eta \phi_{\pm}(x), \hspace{0.25cm} \eta>0.\label{mom1}
\end{eqnarray}
and after a simple integration, we have
\begin{eqnarray}
\phi_{\pm}(x)=\phi_{\pm}(0)\exp\Bigl(\mp\frac{\eta x}{\hbar}\Bigr).
\end{eqnarray}

The operator $\widehat{P}$ is defined on the whole real axis where $\phi_{\pm}(x)$ diverge at $x\rightarrow \pm\infty$ are not normalizable. None of the functions $\phi_{\pm}(x)$ belong to the Hilbert space $\mathcal{L}^{2}(\Re)$. Therefore the deficiency indices are $(0,0)$ and we conclude that the momentum operator is indeed self-adjoint which is similar to the position operator case and with the following domain,
\begin{eqnarray}
\mathcal{D}(\widehat{P})=\mathcal{D}(\widehat{P}^{\dagger})=\{\phi\in\mathcal{D}_{\text{max}}(\widehat{P})\}
\end{eqnarray}  
where $\mathcal{D}_{\text{max}}(\widehat{P})$ denotes the maximum domain in which the operator $\widehat{P}$ has a well defined action, i.e. $\mathcal{D}_{\text{max}}(\widehat{P})=\{\phi\in \mathcal{L}^2(\Re): \widehat{P}\phi\in \mathcal{L}^{2} (\Re)\}$.

The self-adjointness condition of both position and momentum operators in the anti-Snyder model is different as compared to the Snyder model. In the Snyder model, only momentum operator $\widehat{P}$ is self adjoint over the real domain but position operator $\widehat{X}$ is not essentially self-adjoint and merely symmetric \cite{pedramsa, venkat} due to minimal length. However, the position operator in Snyder model has one-parameter family of self-adjoint extension.   

\section{(2+1)-dimensional Relativistic Dirac-Landau Problem with Maximum Momentum}

Following \cite{menculini-landau, menculini-qpt, menculini-qpt-as} we study the relativistic Dirac-Landau problem under anti-Snyder model because the (2+1)-dimensional problem has its importance in various branches of physics, especially condensed matter and nano-physics. This can be seen in graphene where low energy electronic excitations are manifested by massless Dirac fermion with an effective ``light speed" of $v_0 \approx 1.0\times 10^6$ m/s. Also, (2+1)-dimensional systems can act as quantum simulator \cite{bloch} to realize fractional quantum hall effect \cite{stormer} and topological insulators \cite{hasan}. 

From \eqref{MCR2}, for simplicity we consider the (2+1)-dimensional anti-Snyder algebra of the form
\begin{eqnarray}
\left[\widehat{X}_{i}, \widehat{P}_{j}\right] &=& i \hbar \delta_{ij} \bigl(1 - \beta \widehat{P}^2\bigr)\label{MCR3}\\
\left[\widehat{X}_{i}, \widehat{X}_{j}\right] &=& -i \hbar \beta (1-\beta \widehat{P}^2)\widehat{L}_{ij}\label{MCR4}
\end{eqnarray}
where $\widehat{L}_{ij}:=\frac{1}{1-\beta \widehat{P}^2} (\widehat{X}_{i} \widehat{P}_{j}- \widehat{X}_{j} \widehat{P}_{i})$ are the deformed 3-dimensional generators of rotations that act on momentum space wavefunction as
\begin{eqnarray}
\widehat{L}_{ij}\psi(\vec{p}) &=& -i\hbar \Bigl(p_i \frac{\partial}{\partial p_j}- p_j \frac{\partial}{\partial p_i}\Bigr)\psi(\vec{p}).
\end{eqnarray}
To preserve rotational symmetry in the deformed algebra, we require non-commutativity of position operators fundamentally in the theory which is clearly stated in \eqref{MCR4}. Next, we set $i,j=1,2$ for 2-dimensional Dirac-Landau problem and the only orbital angular momentum is    
\begin{eqnarray}
\widehat{L}_{k} &=& \frac{1}{1 - \beta \widehat{P}^2} \epsilon_{ijk} \widehat{X}_{i} \widehat{P}_{j}\nonumber\\
\Rightarrow \widehat{L}_{z} 
&=& \frac{1}{1 - \beta \widehat{P}^2}\bigl(\widehat{X} \widehat{P}_{y}- \widehat{Y} \widehat{P}_{x}\bigr)
=-i \hbar \partial_{\theta} \label{Lz}
\end{eqnarray}
where $(p,\theta)$ is the ``polar coordinate" in momentum space. Furthermore, by requiring symmetricity of the physical operators, we see that the inner product between any two arbitrary wavefunctions \eqref{inner2} becomes
\begin{eqnarray}
\langle \phi|\psi\rangle_f &=& \int_{-\frac{1}{\sqrt{\beta}}}^{\frac{1}{\sqrt{\beta}}} \frac{d^2\vec{p}}{1- \beta p^2} \phi^*(\vec{p})\psi(\vec{p})
= 2\pi\int_{0}^{\frac{1}{\sqrt{\beta}}} \frac{ p\ dp}{1- \beta p^2} \phi^*(\vec{p})\psi(\vec{p})
\end{eqnarray} 
where $p\in[0,1/\sqrt{\beta})$ is the radial momentum and we are only interested in states with rotational symmetry.

Let us consider the (2+1)-dimensional Dirac equation in the presence of a homogeneous magnetic field $\vec{B}= (0, 0, B_0)$. The non-relativistic version of this problem had been solved by Rabi \cite{rabi} back in 1920s. The Hamiltonian of the system can be written as
\begin{eqnarray}
H_D= c\vec{\sigma} \cdot \Bigl(\vec{P} - \frac{q}{c} \vec{A}\Bigr) + \sigma_z Mc^2
\end{eqnarray}
where $q$ is the charged of the fermions\footnote{Subsequently, we will set $q=-e$ for the case of electron.}, $c$ is speed of light and $\vec{\sigma}=(\sigma_x, \sigma_y, \sigma_z)$ are the generators of the internal $SU(2)$ algebra or Pauli matrices. Since $\vec{B}=\nabla \times \vec{A}$ and $\vec{B}$ is homogenous, we have the vector potential in symmetric gauge as
\begin{eqnarray} 
\vec{A} &=& \frac{1}{2} \vec{B}\times \vec{r}=\frac{B_0}{2}\bigl(-y, x, 0\bigr).
\end{eqnarray}

The time independent Dirac equation can be written as,
\begin{eqnarray}
H_D \psi &=& E \psi; \hspace{0.5cm} \psi
= \begin{pmatrix}
  \psi^{(1)} \\
  \psi^{(2)} 
 \end{pmatrix} \label{diracH1}
\end{eqnarray}
where $\psi=(\psi^{(1)}, \psi^{(2)})^{t}$ is the two-components Dirac spinors and $``t"$ refers to transposition operation. By using representation \eqref{rep.P} and the Pauli matrices, we have
\begin{eqnarray}
H_D \psi &=& \left[c \begin{pmatrix}
  0 & P_x - i P_y\\
   P_x + i P_y & 0
 \end{pmatrix} + e \begin{pmatrix}
  0 & A_x - i A_y\\
   A_x + i A_y & 0
 \end{pmatrix} + \begin{pmatrix}
  M c^2 & 0\\
   & -M c^2
 \end{pmatrix} \right] \psi \nonumber\\
&=&\begin{pmatrix}
  M c^2 & c P_{-}\\
   c P_{+} & -M c^2
 \end{pmatrix} \begin{pmatrix}
  \psi^{(1)} \\
  \psi^{(2)} 
 \end{pmatrix} = E \begin{pmatrix}
  \psi^{(1)} \\
  \psi^{(2)} 
 \end{pmatrix} \label{diracH2}
\end{eqnarray}
where it is useful to define the ladder operators
\begin{eqnarray}
P_{\pm} &=& \bigl(P_x + \frac{e}{c}\ A_x\bigr) \pm i \bigl(P_y + \frac{e}{c}\ A_y\bigr). \label{ladder1} 
\end{eqnarray}
By denoting the following quantities,
\begin{eqnarray}
p_x &=& p\cos\theta, \ \ p_y = p\sin\theta, \ \ p^2 = p^2_x + p^2_y \\  
\lambda &=& \frac{e\hbar B_0}{2c}, \hspace{0.5cm} \bigl[\lambda\bigr]=\bigl[\text{momentum}^2\bigr]
\end{eqnarray}
the ladder operators \eqref{ladder1} can be rewritten in momentum representation
\begin{eqnarray}
P_{\pm} &=& p e^{\pm i\theta} -\lambda (1-\beta p^2) \bigl(\pm \partial_{p_{x}} + i \partial_{p_{y}} \bigr)\nonumber\\
&=& p e^{\pm i\theta} -\lambda (1-\beta p^2) \left[\pm \Bigl(\frac{p\cos\theta}{p}\partial_p-\frac{p\sin\theta}{p^2}\partial_{\theta}\Bigr)+ i \Bigl(\frac{p\sin\theta}{p}\partial_p +\frac{\cos\theta}{p}\partial_{\theta}\Bigr)\right]\nonumber\\
&=& e^{\pm i\theta}\Bigl[p -\lambda (1-\beta p^2) \Bigl(\pm \partial_p + \frac{i}{p}\partial_{\theta}\Bigr)\Bigr].
\end{eqnarray}
They satisfy the usual ladder operators relations $[L_{z}, P_{\pm}]=\pm P_{\pm}$ and $[P_{+}, P_{-}]= 2\hbar L_{z}$. This further confirms that $P_{\pm}$ are indeed the ladder operators. Including spin, the total angular momentum operator is $J_{z} = L_{z}+ (\hbar/2)\sigma_{z}$ which commute with the Dirac Hamiltonian, i.e. $[H_{D}, J_{z}]=0$. Thus, the total angular momentum is conserved even in the new anti-Snyder algebra with the eigenvalue of $J_{z}$ as $j=\hbar (m \pm 1/2)$. Here symbol $``m"$ carries the usual meaning of orbital angular momentum quantum number.

We can decompose \eqref{diracH2} into the components where they are coupled through
\begin{eqnarray}
P_{+}\psi^{(1)} &=& \epsilon_{+} \psi^{(2)}\ ; \hspace{0.5cm} P_{-}\psi^{(2)} = \epsilon_{-} \psi^{(1)}\\
\epsilon_{\pm} &:=& \frac{E\pm M c^2}{c}.
\end{eqnarray}
These coupled equations can be decoupled into two independent equations,
\begin{eqnarray}
P_{-} P_{+} \psi^{(1)} &=& P_{-} \epsilon_{+} \psi^{(2)} = \epsilon^2 \psi^{(1)} \label{intertw1}\\
P_{+} P_{-} \psi^{(2)} &=& P_{+} \epsilon_{-} \psi^{(1)} = \epsilon^2 \psi^{(2)} \label{intertw2}\\
\epsilon^2 &:=& \frac{E^2- M^2 c^4}{c^2}.
\end{eqnarray}

Since $\{H_D, J_{z}\}$ form a set of commuting observables and simultaneous eigenstates, we can use the eigenstates of the total angular momentum to represent the wavefunctions of the problem by putting them in the form,
\begin{eqnarray}
\psi_m^{(1)} &=& u_m^{(1)}(p) e^{i m \theta}, \hspace{0.5cm} \psi_m^{(2)} = u_m^{(2)}(p) e^{i (m+1) \theta}. 
\end{eqnarray}  
Using the above ansatz on the wavefunctions, the decoupled equations in the momentum representation read,
\begin{eqnarray}
&& P_{\mp} P_{\pm} \psi_{m}^{(j)} = P_{\mp} P_{\pm} u_m^{(j)}(p) e^{i m' \theta}\nonumber\\
&& = P_{\mp}\left(e^{\pm i\theta}\Bigl[p\ \psi_{m}^{(j)}(p) -\lambda (1-\beta p^2)\Bigl(\pm e^{i m' \theta}\partial_{p} u_{m}^{(j)}(p) -\frac{m'}{p}\psi_m^{(j)}(p)\Bigr)\Bigr]\right)\nonumber\\
&&= e^{\mp i\theta}\Bigl[p -\lambda (1-\beta p^2) \Bigl(\mp \partial_p + \frac{i}{p}\partial_{\theta}\Bigr)\Bigr]\nonumber\\
&&\phantom{|}\times\left(e^{\pm i\theta}\Bigl[p\ \psi_{m}^{(j)}(p) -\lambda (1-\beta p^2)\Bigl(\pm e^{i m' \theta}\partial_{p} u_{m}^{(j)}(p) -\frac{m'}{p}\psi_m^{(j)}(p)\Bigr)\Bigr]\right)\nonumber\\
&&= \epsilon^2 \psi_{m}^{(j)},
\end{eqnarray}
where we have denoted $m'= m$ and $m'=m+1$ for $j=1,2$ respectively. Simplifying, it reduces to
\begin{eqnarray}
&&\left\{ p^2 + 2\lambda(1-\beta p^2)\left[m'\pm 1 + \lambda \beta\Bigl(p \frac{d}{dp} \mp m'\Bigr)\right]\right.\nonumber\\
&& \left.-\lambda^2 (1-\beta p^2)^2 \Bigl(\frac{d^2}{dp^2} + \frac{1}{p} \frac{d}{dp} - \frac{m'^2}{p^2} \Bigr)\right\} u_m^{(j)} (p) = \epsilon^2 u_m^{(j)} (p). \label{diraceqn}
\end{eqnarray} 

To solve \eqref{diraceqn} we perform a change of variable in two steps. Firstly, by letting $u_m^{(j)}= p^{-1/2} \varphi_m^{(j)}$, \eqref{diraceqn} becomes
\begin{eqnarray}
&&\left\{ p^2 + 2\lambda(1-\beta p^2)\left[m'\pm 1 + \lambda \beta\Bigl(p \frac{d}{dp} \mp m' -\frac{1}{2}\Bigr)\right]\right.\nonumber\\
&& \left.-\lambda^2 (1-\beta p^2)^2 \Bigl(\frac{d^2}{dp^2} + \frac{1}{p^2} \Bigl(\frac{1}{4}- m'^2\Bigr)\Bigr)\right\} \varphi_m^{(j)} (p) = \epsilon^2 \varphi_m^{(j)} (p). \label{diraceqn1}
\end{eqnarray} 
In order to cast the above equation into Schr$\ddot{\text{o}}$dinger-like equation, we perform another change of variables\footnote{The choice of variable results from demanding the resulting differential equation does not contain first order derivative. It can be found by letting $p=\frac{f(q)}{\sqrt{\beta}}$ then requiring the coefficient of the first order derivative to vanish.}
\begin{eqnarray}
p &=& \frac{\tanh q}{\sqrt{\beta}}, \hspace{0.25cm} q\in (0, +\infty).
\end{eqnarray}
Note that new variable ensures that $0 < p < 1/\sqrt{\beta}$ since $\tanh q \rightarrow 1$ as $q\rightarrow \infty$. 


In the $q$ domain,  \eqref{diraceqn} takes the form of
\begin{eqnarray}
&&\left\{-\lambda^2 \beta \frac{d^2}{dq^2} +\frac{2\lambda}{\cosh^2 q}\Bigl[(m' \pm 1) + \lambda \beta (\mp m'-1/2)\Bigr] + \frac{\tanh^2 q}{\beta}\right.\nonumber\\
&& \left.-\frac{\lambda^2 \beta}{\sinh^2 q \cosh^2 q}\Bigl(1/4 -m'^2\Bigr)\right\} \varphi_m^{(j)} (q) = \epsilon^2 \varphi_m^{(j)} (q). \label{diraceqn2}
\end{eqnarray} 
This is the Schr$\ddot{\text{o}}$dinger-like equation with some complicated hyperbolic potentials. We simplify the potential term by complexifying the radial momentum through the following transformation
\begin{eqnarray}
q &=& \frac{x}{2} + i \frac{\pi}{4} , \hspace{0.25cm} x\in [0, +\infty].
\end{eqnarray}
where the real $p$ can be obtained through 
\begin{eqnarray}
p=\text{Re} \Bigl[\frac{\tanh(x/2 + i \pi/4)}{\sqrt{\beta}}\Bigr].
\end{eqnarray}

With these transformations, Eq.\eqref{diraceqn2} in the $x$ variable reduces to 
\begin{eqnarray}
\left(-\frac{d^2}{d x^2} + \mu\ \text{sech}^2 x+ i \nu\ \text{sech} x \tanh x\right) \varphi_m^{(j)} (x) &=& k^2 \varphi_m^{(j)} (x) \label{diraceqn3}\\
\Rightarrow\ \left(-\frac{d^2}{d x^2} + V^{(\mu, \nu)}_{\text{hs}} (x)\right) \varphi_m^{(j)} (x) &=& k^2 \varphi_m^{(j)} (x) \label{diraceqn4}
\end{eqnarray}
where we have defined the dimensionless quantities,
\begin{eqnarray}
\mu &=&\frac{1}{\beta\lambda}\Bigl[m' \pm 1 +\beta \lambda \bigl(\mp m' -1/2\bigr)\Bigr] +\frac{1}{4} -m'^2 - \frac{1}{2\beta^2 \lambda^2}\\
\nu &=& \frac{1}{2\beta^2 \lambda^2} -\frac{1}{\beta \lambda}\Bigl[m' \pm 1 +\beta \lambda \bigl( \mp m' -1/2\bigr)\Bigr]\\
k^2 &=& \frac{\epsilon^2 - 1/\beta}{4\beta \lambda^2}
\end{eqnarray}
and the complexified hyperbolic Scarf potential (CHSP) \cite{znojil}
\begin{eqnarray}
V^{(\mu, \nu)}_{\text{hs}}(x) &=& \mu\ \text{sech}^2 x + i \nu\ \text{sech} x \tanh x, \hspace{0.3cm} x\in [0, +\infty]. \label{hyperscarf}
\end{eqnarray}
Here, we assume $\mu <0 \ ; \ \nu\neq 0$. We see that \eqref{hyperscarf} exhibits a shape invariance and $\mathcal{PT}-$symmetry properties in the context of Bender-Boettcher non-Hermitian quantum mechanics \cite{nhqm}. Here $\mathcal{P}$ changes the parity $\tilde{x}\rightarrow -\tilde{x}$ and the complex conjugation operator $\mathcal{T}$ transforms $i$ into $-i$ hence mimics the time reversal symmetry. $\mathcal{PT}$-symmetry often leads the complex potential to the fully real and discrete spectrum through the pseudo-hermiticity condition \cite{pseudoh}. Eq.\eqref{hyperscarf} is an exactly solvable model defined by
\begin{eqnarray}
V^{(A,B)}_{\text{hs}} (x)= - \bigl(B^2 + A (A+1)\bigr)\ \text{sech}^2 x+ i B (2A+1)\ \text{sech} x \tanh x \label{hyperscarf1}
\end{eqnarray}
where $A, B \in \Re$. The real and imaginary parts of the potential have no singularity on the real axis. Interestingly, CHSP \eqref{hyperscarf1} is invariant under the exchange of $A + \frac{1}{2}\ \Leftrightarrow\ B$ so usually we can assume $A + \frac{1}{2} >0 ; B>0$ without loss of generality \cite{bagchi}. However, in this article we shall let both $A,B$ to be real-valued function of angular momentum, $m$. The only constraint on them will be the requirements from the physical aspects, i.e. normalizability and orthogonality conditions of the wavefunctions etc. that we will discuss in next section. This CHSP \eqref{hyperscarf1} has been found independently within the framework of supersymmetric quantum mechanics with shape invariant super-potential \cite{susyqm, susycomplexp} of the form,
\begin{eqnarray}
W^{(A,B)}_{\text{hs}}(x) = A \tanh x + i B \text{sech}x \label{hyperscarf2}
\end{eqnarray}
with real eigenspectra 
\begin{eqnarray}
E_{n}= - (A-n)^2; \hspace{0.25cm} n_{\text{max}} < A. \label{hsE}
\end{eqnarray}
There is an upper bound for the maximum number of bound states. We interpret this as the genuine effects induced by MCR with maximum momentum cutoff, see \cite{mig1, jiz, cutoff, pedrammax, potwell, spectra, coherent, rwe}.

As a consequence, we can recast the Schr$\ddot{\text{o}}$dinger-like equation \eqref{diraceqn4} into
\begin{eqnarray}
\left(-\frac{d^2}{d x^2} + V^{(A, B)}_{\text{hs}}(x)\right) \varphi_m^{(j)} (x) &=& k^2 \varphi_m^{(j)} (x) \label{diraceqn5}
\end{eqnarray}
with the identification of
\begin{eqnarray}
\mu &=& - (B^2 + A(A+1)) \nonumber\\
&=& \frac{1}{\beta\lambda}\Bigl[m' \pm 1 +\beta \lambda \bigl( \mp m' -1/2\bigr)\Bigr] +\frac{1}{4} -m'^2 - \frac{1}{2\beta^2 \lambda^2} < 0 \label{cond1}\\
\nu &=& B (2A+1) \nonumber\\
&=& \frac{1}{2\beta^2 \lambda^2} -\frac{1}{\beta \lambda}\Bigl[m' \pm 1 +\beta \lambda \bigl( \mp m' -1/2\bigr)\Bigr] \neq 0 \label{cond2}
\end{eqnarray}
which we have set $B^2 > -A(A+1);$\ $A\neq -\frac{1}{2}$ and $B\neq 0$. We need to solve the constants $A$ and $B$ simultaneously. The physical solutions are classified by them in the next section. After solving\footnote{Recall that $m'=m$ for $j=1$ and $m'=m+1$ for $j=2$. }, we have for upper component of wavefunction $\varphi_m^{(j=1)}$,

\begin{eqnarray}
A &=& m -\frac{1}{2\beta \lambda}, \hspace{1.65cm} B=\frac{1}{2}  -\frac{1}{2 \beta \lambda}\label{a1}\\
A &=& -1 - m + \frac{1}{2\beta \lambda}, \hspace{0.75cm} B=-\frac{1}{2}  +\frac{1}{2 \beta \lambda}\label{a2}\\
A &=& -\frac{1}{2\beta \lambda}, \hspace{2.25cm} B=\frac{1}{2} +m -\frac{1}{2 \beta \lambda}\label{a3}\\
A &=& -1 + \frac{1}{2\beta \lambda}, \hspace{1.6cm} B=-\frac{1}{2} - m  +\frac{1}{2 \beta \lambda}\label{a4}
\end{eqnarray}
while for the lower components $\varphi_m^{(j=2)}$,
\begin{eqnarray}
A &=& m -\frac{1}{2\beta \lambda}, \hspace{1.65cm} B=-\frac{1}{2}  -\frac{1}{2 \beta \lambda}\label{a5}\\
A &=& -1 - m + \frac{1}{2\beta \lambda}, \hspace{0.75cm} B=\frac{1}{2}  +\frac{1}{2 \beta \lambda}\label{a6}\\
A &=& \frac{1}{2\beta \lambda}, \hspace{2.5cm} B=-\frac{1}{2} - m +\frac{1}{2 \beta \lambda}\label{a7}\\
A &=& -1 - \frac{1}{2\beta \lambda}, \hspace{1.6cm} B=\frac{1}{2} + m  -\frac{1}{2 \beta \lambda}\label{a8}.
\end{eqnarray}
The finite orthogonality property of the wavefunctions will impose bounds on the values of parameter $A$ and $B$.

\subsection{Results and Discussions}

The CHSP in Eq.\eqref{diraceqn5} can be further reduced to a hypergeometric equation \cite{alvarez1} and the solutions are the finite Romanovski polynomials \cite{romanovski} denoted as $R_n^{(a+1/2, -2ib)} (z)$. They can be constructed through the Rodrigues representations. See appendix for details. The eigenfunctions of \eqref{diraceqn5} are readily obtained from \cite{alvarez1, alvarez2},
\begin{eqnarray}
\varphi_n^{(j)} (z=\sinh^{-1}x) = (1+ x^2)^{-\frac{A}{2}} e^{-i B \tan^{-1}x} R_n^{(A+\frac{1}{2}, -2iB)} (x) \label{eigenfn}
\end{eqnarray}
where $x \in (0 < x = \sinh z <+\infty)$. Also, the eigenenergies are given by
\begin{eqnarray}
k^2 &=& - (A-n)^2\nonumber\\
\Rightarrow E_{n}^2 &=& M^2 c^4 + c^2 \Bigr( \frac{1}{\beta} - 4\beta \lambda^2 (A-n)^2 \Bigr) \label{eigenE}. 
\end{eqnarray}
Both $A$ and $B$ are parameters of the Romanovski polynomial, which are solved through the simultaneous equation \eqref{cond1} and \eqref{cond2} and listed in Eq.\eqref{a1} $-$ Eq.\eqref{a8}. $n$ is the order of the polynomial, hence $n=0, 1, 2...n_{\text{max}} <A $.  

The non-trivial orthogonality condition of the Romanovski polynomial is given by
\begin{eqnarray}
\int_{-\infty}^{\infty}dz\ \varphi_n^{(j)} (z) \varphi_{n'}^{(j)} (z) &=&\int_{-\infty}^{\infty}dx  (1+ x^2)^{-(A+\frac{1}{2})} e^{-2i B \tan^{-1}x} \nonumber\\
&&\hspace{0.5cm}\times R_n^{(A+\frac{1}{2}, -2iB)} (x) R_{n'}^{(A+\frac{1}{2}, -2iB)} (x) \label {orthc}
\end{eqnarray}
where we have the weight function
\begin{eqnarray}
w^{(A+\frac{1}{2}, -2iB)} (x) = (1+ x^2)^{-(A+\frac{1}{2})} e^{-2i B \tan^{-1}x}. \label{wgt}
\end{eqnarray}
As long as the weight function decreases as $x^{-2(A+\frac{1}{2})}$, integral of Eq.\eqref{orthc} will be convergent only if 
\begin{eqnarray}
n + n' < 2\bigl(A+1/2\bigr) -1 = 2A.
\end{eqnarray}
This means that only a finite number of Romanovski polynomials are orthogonal to each other. These Romanovski polynomials serve as the wavefunctions for CHSP. We have a finite number of bound states determined by the inequality, where the uppermost bound state is bounded by the value of $A$, that is $n_{\text{max}} < A$. The fall-off condition of weight function imposes a constraint on the constant 
\begin{eqnarray}
&& A+ \frac{1}{2}> 0 \Rightarrow A > -\frac{1}{2}.\label{cond3}
\end{eqnarray}
Also, we expect the modified energy spectrum should be equal or greater than the rest energy of the electrons, i.e. $E_n^2 \geq M^2c^4$. Setting $n=0$, we obtain another constraint on the constant $A$
\begin{eqnarray}
&&\frac{1}{\beta^2} - 4\beta \lambda^2 A^2 \geq 0 \Rightarrow\ A\leq \frac{1}{2\beta \lambda}.\label{cond4}
\end{eqnarray}
Thus, from both Eq.\eqref{cond3}, \eqref{cond4} and assumption on values of $A$ and $B$ we introduced in \eqref{cond1} and \eqref{cond2}, we have
\begin{eqnarray}
\boxed{-\frac{1}{2} < A \leq  \frac{1}{2\beta \lambda};\hspace{0.25cm} A\neq -\frac{1}{2}; \hspace{0.25cm} B\neq 0\label{cond5}}
\end{eqnarray} 
to be satisfied by acceptable parameters $A$ and $B$. Note that the deformation parameter $\beta \geq 0$ and constant $\lambda>0$. By direct comparison, we see that the value of $A$ in \eqref{a8} do not satisfy the requirement imposed by \eqref{cond5} for all value of $\frac{1}{\beta\lambda}$ and thus cannot be part of the solutions. Also, solution \eqref{a3} is only valid for $\frac{1}{\beta\lambda} < 1$. Meanwhile, all the constant $B$'s are also required to be non-vanishing. 

Using Eq.\eqref{cond5} we can classify the eigenfunctions and corresponding eigenenergy spectrum according to the angular momentum quantum number $m$. The eigen-solutions are arranged according to the orbital angular quantum number $m$ and we build the spinors solution by putting together those states of $\psi^{(1)}$ and $\psi^{(2)}$ which have the same energy eigenvalues\footnote{We only consider positive branch of the spectrum and the negative branch can be obtained through Eq.\eqref{eigenE} by the negative root solutions.}. It is crucial to emphasize that the spinors formed have to satisfy the intertwining condition in eq.\eqref{intertw1}, \eqref{intertw2}. We denote
\begin{eqnarray}
\psi_{m,n}^{(1)} (p) &=& \frac{e^{-i B \text{tan}^{-1} \frac{\sqrt{\beta}p}{\sqrt{1-\beta p^2}}}}{\sqrt{p(1-\beta p^2)^A}}\ R_n^{(A+\frac{1}{2},-2i B)}\Bigl(\frac{\sqrt{\beta}p}{\sqrt{1-\beta p^2}}\Bigr)\ e^{i m \theta}\label{psi1}\\
\psi_{m,n}^{(2)} (p) &=& \frac{e^{-i B \text{tan}^{-1} \frac{\sqrt{\beta}p}{\sqrt{1-\beta p^2}}}}{\sqrt{p(1-\beta p^2)^A}} \ R_n^{(A+\frac{1}{2},-2i B)}\Bigl(\frac{\sqrt{\beta}p}{\sqrt{1-\beta p^2}}\Bigr)\  e^{i (m+1)\theta} \label{psi2}.
\end{eqnarray}
The solutions in which the energy levels are dependent of the orbital angular momentum $m$ are shown in table 1 and table 2. Next, for the cases in which energy levels are independent of $m$, we divide the solutions into the cases $\frac{1}{\beta\lambda} > 1$ (table 3) and $\frac{1}{\beta\lambda} \leq 1$ (table 4) accordingly.

\begin{table}[ht]
\caption{Energy levels and the corresponding wavefunctions for $-\frac{1}{2} + \frac{1}{2\beta\lambda} < m \leq \frac{1}{\beta\lambda}$. From $A\neq -\frac{1}{2}\Rightarrow m\neq -\frac{1}{2} + \frac{1}{2\beta\lambda}$. Also, $B\neq 0 \Rightarrow \frac{1}{\beta\lambda} \neq 1$. Denote $N_{\pm}= m\pm n$.}
\centering
\begin{tabular}{c c c}
\hline\hline
$E_{m,n} = \sqrt{M^2 c^4 + 2\hbar e B_0 c N_{-} \Bigl[1 - \beta \frac{\hbar e B_0}{2c}N_{-}\Bigr]}$, & $\psi_{m,n}(p) =
 \begin{pmatrix}
 \psi^{(1)}_{m,n}(p) \\
 \psi^{(2)}_{m,n}(p)
 \end{pmatrix}$ ,
& $n=0,1,2, ... < A$ \\ 
$\psi_{m,n}^{(1)} (p)$ & $A=m-\frac{1}{2\beta\lambda}$\ , & $B=\frac{1}{2} \bigl(1-\frac{1}{\beta\lambda}\bigr)$\\
$\psi_{m,n}^{(2)} (p)$ & $A=m-\frac{1}{2\beta\lambda}$\ , & $B=\frac{1}{2} \bigl(-1-\frac{1}{\beta\lambda}\bigr)$\\ [1ex] 
\hline\hline
\end{tabular}
\label{table:I}
\end{table}

\begin{table}[ht]
\caption{Energy levels and the corresponding wavefunctions for $-1 \leq m < -\frac{1}{2} + \frac{1}{2\beta\lambda}$. Similarly to table (1), $m\neq -\frac{1}{2} + \frac{1}{2\beta\lambda}$ and $\frac{1}{\beta\lambda} \neq 1$. Denote $N_{\pm}= m\pm n$.}
\centering
\begin{tabular}{c c c}
\hline\hline
$E_{m,n} = \sqrt{M^2 c^4 + 2\hbar e B_0 c(N_{+} +1) \Bigl[1-\beta \frac{\hbar e B_0}{2c}(N_{+} +1)\Bigr]}$, & 
$\psi_{m,n}(p) =
\begin{pmatrix}
\psi^{(1)}_{m,n}(p) \\
\psi^{(2)}_{m,n}(p)
\end{pmatrix}$,
& $n=0,1,2, ...< A$ \\
$\psi_{m,n}^{(1)} (p)$ & $A= -1 -m + \frac{1}{2\beta\lambda}$\ , & $B=\frac{1}{2} \bigl(-1+\frac{1}{\beta\lambda}\bigr)$ \\
$\psi_{m,n}^{(2)} (p)$ & $A= -1 -m + \frac{1}{2\beta\lambda}$\ , & $B=\frac{1}{2} \bigl(1+\frac{1}{\beta\lambda}\bigr)$ \\ [1ex] 
\hline\hline
\end{tabular}
\label{table:II}
\end{table}

\begin{table}[ht]
\caption{Energy levels and the corresponding wavefunctions which is independent of $m$. Here we assume the case with $\frac{1}{\beta\lambda} > 1$. However, $A\neq -\frac{1}{2}\Rightarrow \frac{1}{\beta\lambda}\neq \pm 1$ which is trivially valid. Also, $B\neq 0\Rightarrow m\neq -\frac{1}{2} + \frac{1}{2\beta\lambda}$.}
\centering
\begin{tabular}{c c c}
\hline\hline
$E_{m,n} = \sqrt{M^2 c^4 + 2\hbar e B_0 c n \Bigl[1-\beta \frac{\hbar e B_0}{2c} n \Bigr]}$, & $\psi_{m,n}(p) =\begin{pmatrix}
\psi^{(1)}_{m,n-1}(p) \\
\psi^{(2)}_{m,n}(p)
\end{pmatrix}$, & 
$n = 1,2, ... < \frac{1}{2\beta\lambda}$ \\ 
$\psi_{m,n-1}^{(1)} (p)$ & $A= -1 + \frac{1}{2\beta\lambda}$\ , & $B=\frac{1}{2} \bigl(-1 -2m +\frac{1}{\beta\lambda}\bigr)$\\
$\psi_{m,n}^{(2)} (p)$ & $A=\frac{1}{2\beta\lambda}$\ , & $B=\frac{1}{2} \bigl(-1 - 2m +\frac{1}{\beta\lambda}\bigr)$\\ [1ex] 
\hline
$E_{m,0} = M c^2$, & $\psi_{m,0}(p) =
\begin{pmatrix}
0 \\
\psi^{(2)}_{m,0}(p)
\end{pmatrix},$ & $n=0$\\
$\psi_{m,0}^{(2)} (p)$ & $A = \frac{1}{2\beta\lambda}$\ , & $B = \frac{1}{2} \bigl(-1 -2m +\frac{1}{\beta\lambda}\bigr)$\\ [1ex] 
\hline\hline
\end{tabular}
\label{table:III}
\end{table}

\begin{table}[ht]
\caption{Energy levels and the corresponding wavefunctions which is independent of $m$. Here we assume the case with $\frac{1}{\beta\lambda} \leq 1$. $A\neq -\frac{1}{2}$ is trivially satisfied. Also, $B\neq 0\Rightarrow m\neq -\frac{1}{2} + \frac{1}{2\beta\lambda}$.}
\centering
\begin{tabular}{c c c}
\hline\hline
$E_{m,n} = \sqrt{M^2 c^4 + 2\hbar e B_0 c n \Bigl[1-\beta \frac{\hbar e B_0}{2c} n \Bigr]}$, & $\psi_{m,n}(p) =\begin{pmatrix}
0 \\
\psi^{(2)}_{m,n}(p)
\end{pmatrix},$ & 
$n = 0, 1,2, ... < \frac{1}{2\beta\lambda}$ \\
$\psi_{m,n}^{(2)} (p)$ & $A=\frac{1}{2\beta\lambda}$\ , & $B=\frac{1}{2} \bigl(-1 - 2m +\frac{1}{\beta\lambda}\bigr)$\\ [1ex]\hline\hline
\end{tabular}
\label{table:IV}
\end{table}

\newpage
The above tables show that, under the anti-Snyder's algebra, the energy spectrum of (2+1)-dimensional Dirac equation in a homogenous magnetic field results in three distinct bands characterized by angular momentum $m$ in contrast to the usual case. Among them, two out of three bands (table 1 and table 2) have range of values of angular momentum quantum number $m$ which are dependent on the value of the deformation parameter $\beta$. In general, the eigenenergies are shifted negatively compared to the undeformed spectrum. This feature is the generic phenomena due to the maximum momentum cutoff in the anti-Synder model and it has been observed in other deformed quantum mechanical system \cite{cutoff, jiz, pedrammax, spectra, potwell, coherent, rwe}. This is in contrast to the Snyder's algebra which encodes a minimal length\cite{menculini-landau}. The eigenenergies of deformed relativistic Dirac-Landau problem in Snyder's model is shifted positively compared to the undeformed case\cite{menculini-landau, menculini-qpt}. 



Table 1 gives the energies and eigenfunctions with angular momentum quantum number $-\frac{1}{2} + \frac{1}{2\beta\lambda} < m \leq \frac{1}{\beta\lambda}$. The maximum number of principle quantum number is given by $n_{\text{max}} < A = m- \frac{1}{2\beta\lambda}$. In this band of solution $N_{-} = (m-n)>0$, the eigenenergies are negatively deformed. This upper bound on the total number of allowed bound states means there is an upper bound on the energy which implies a finite maximum momentum. In the limit $\beta\rightarrow 0$, all the states in this band vanish as the range of $m$ becomes physically meaningless. We interpret this as the genuine effects that deviate from ordinary quantum mechanics due to the new deformed anti-Snyder algebra.  

Table 2 shows the solutions for values of the angular momentum quantum number in the range $-1 \leq m < -\frac{1}{2} + \frac{1}{2\beta\lambda}$. Within this band all energy levels except lowest state have a finite degeneracy. For instance, for $ n + m = N_{+}$, such that $n_{\text{max}} + m < \left[\frac{1}{2\beta\lambda} -1\right]$ where $[x]$ represents the smallest integer equal or less than $x$, the levels are $D = (N_{+} + 2)$-fold degenerate. The ground state energy is unique and is given by $E_{(-1,0)} = M c^2$ when $n + m = N_{+} = -1$. All the states are spin doublet, i.e. they have a spin-up $\psi^{(1)}$ as well as a corresponding spin down $\psi^{(2)}$ component. In the limit $\beta\rightarrow 0$, the range of possible angular quantum number becomes $-1 \leq m < \infty$. Hence this class of solutions reduces to the ordinary solutions of relativistic Dirac-Landau problem with $m=-1, 0 $ and all positive values of $m$. The degeneracy which is proportional to $m$ shall tends to infinite in this limit.

Table 3 gives all solutions valid for all $m \bigl(m\neq -\frac{1}{2} + \frac{1}{2\beta\lambda}\bigr)$ with $\frac{1}{\beta\lambda} >1$. The ground state is a spin singlet whereas all the excited states are spin doublets. Since the energies of this class of solutions are independent of angular momentum quantum number $m$\footnote{Here $m$ is not subjected to finite bound}, we do not have finite degeneracy. In the limit $\beta\rightarrow 0$, the solutions are valid for $m\in (-\infty, \infty)$, which reduces to the ordinary case.

Table 4 gives all solutions valid for all $m \bigl(m\neq -\frac{1}{2} + \frac{1}{2\beta\lambda}\bigr)$ with $\frac{1}{\beta\lambda} \leq 1$. Unlike solutions in table 3, here all the valid solutions are spin singlet. Note that
\begin{eqnarray}
\frac{1}{2\beta\lambda} &=& \frac{c}{\beta e \hbar B_{0}} = \Bigl(\frac{1}{\sqrt{\beta}}\Bigr)^2 \cdot\frac{1}{\hbar M} \cdot\frac{Mc}{e B_0}\nonumber\\
&=& \frac{p_{\text{max}}^2}{\hbar^2} \cdot\frac{\hbar}{M\omega_{L}} = \Bigl(\frac{p_{\text{max}}\ l_{c} }{\hbar}\Bigr)^2 \gg 1\label{scale}
\end{eqnarray}
where $p_{\text{max}}$ is the maximum momentum cutoff imposed by the model, $\omega_{L}=\frac{e B_0}{Mc}$ denotes the electron cyclotron frequency and $l_c=\sqrt{\frac{\hbar}{M \omega_{L}}}$ is the characteristic length scale of the corresponding oscillator. From phenomenological point of view, since the standard quantum mechanics has been tested to very high accuracy, we aspect the new deformation parameter (assume to be universal here) is very small in nature, that is $\beta\ll 1$, thus $p_{\text{max}}=\frac{1}{\sqrt{\beta}}\gg 1$. Hence, the solution in table 4 plays minimum importance from the phenomenological point of view.

\subsection{Massless Dirac-Landau Problem in (2+1) Dimensions}

The Hamiltonian for electrons in graphene in the presence of perpendicular magnetic field is effectively similar to the (2+1)-dimensional massless Dirac equation with Fermi velocity $v_F \approx \frac{c}{300}$,
\begin{eqnarray}
H_{D, M=0} &=& c \vec{\sigma} \cdot \bigl(\vec{P} + \frac{e}{c} \vec{A}\bigr) \Rightarrow H_{\text{graphene}} = v_{F} \vec{\sigma} \cdot \bigl(\vec{P} + \frac{e}{c} \vec{A}\bigr).
\end{eqnarray} 

Following the discussion in Sect.(3) to solve this problem, only some minor changes are required, which is by setting $c=v_{F}, M=0$ and thus
\begin{eqnarray}
\epsilon \rightarrow \epsilon' = \frac{E}{v_F}.
\end{eqnarray}
This means that the above discussions are similar except a minor change to the energy levels,
\begin{eqnarray}
E_{n,m} &=& v_F \sqrt{\frac{1}{\beta} - 4\beta \lambda^2 (A-n)^2} \label{eigenEm=0}.
\end{eqnarray}
Subsequently, we present the eigenenergies spectrum in table 5. In the limit $\beta\rightarrow 0$, the solutions in the first and the last row are no longer valid. Second and third solutions reproduce the eigenstates of the standard quantum mechanics respectively. Consider the energy spectrum for the case of small deformation i.e. $\beta \ll 1$ which is phenomenologically favoured. Expanding the third energy level in table 5 in terms of $\beta$, we get
\begin{eqnarray}
E_{n} &=& v_F \sqrt{\frac{2\hbar e B_0 n}{c}} \left(1-\beta \frac{\hbar e B_0}{2c} n \right)^{1/2}\label{eigenE1}\\
&=& v_F \sqrt{\frac{2\hbar e B_0 n}{c}} - \beta v_F \Big(\frac{\hbar e B_0 n}{2c}\Bigr)^{3/2} + O(\beta^2). \label{eigenE2}
\end{eqnarray}   
We see that the leading order energy shifts of the eigenstates are of opposite sign compared to those obtained from string theory motivated MCR’s, e.g. Snyder model case \cite{menculini-landau}. Thus these two cases are easily distinguishable in experiments involving measurement of Landau levels transitions, for example the infrared spectroscopy study of transitions between graphene Landau levels \cite{zjiang}. In Fig.(1), we plot the energies \eqref{eigenE1} and show the deviation for the both Snyder/anti-Snyder case with respect to the undeformed case. We see that for finite $\beta\neq 0$, there is a maximum allowed bound state in anti-Snyder model, but not in the Snyder model with minimal length.  

\begin{table}[ht]
\caption{Energy levels for massless electrons. The degeneracy of the energy levels are similar to the massive case. Denote $N_{\pm}=m\pm n$ and $\epsilon_{n,m} = \frac{E_{n,m}}{v_{F} \sqrt{2\hbar e B_0/c}}$.}
\centering
\begin{tabular}{c c c}
\hline\hline
$-\frac{1}{2} + \frac{1}{2\beta\lambda} < m \leq \frac{1}{\beta\lambda}$\ , & $n=0,1,... < \left[m-\frac{1}{2\beta\lambda}\right]$\ ,
& $\epsilon_{n,m}= N_{-}^{\frac{1}{2}} \Bigl[1 - \beta \frac{\hbar e B_0 N_{-}}{2c} \Bigr]$\\ [1ex] \hline
$-1 \leq m < -\frac{1}{2} + \frac{1}{2\beta\lambda}$ & $n=0,1,... < \left[-1-m+\frac{1}{2\beta\lambda}\right]$, & $\epsilon_{n,m} = (N_{+} +1)^{\frac{1}{2}}\Bigl[1-\beta \frac{\hbar e B_0}{2c}(N_{+}+1)\Bigr]$ \\ [1ex] \hline
$m\neq -\frac{1}{2} + \frac{1}{2\beta\lambda}$\ ; \ $\frac{1}{\beta\lambda} >1$ & $n = 0$ & $E_0 = 0$ \\ & $n = 1,2, ... < \left[\frac{1}{2\beta\lambda}\right]$ & $\epsilon_{n,0} = \sqrt{n} \Bigl[1-\beta \frac{\hbar e B_0}{2c} n \Bigr]^{\frac{1}{2}}$\\ [1ex] \hline
$m\neq -\frac{1}{2} + \frac{1}{2\beta\lambda}$\ ; \ $\frac{1}{\beta\lambda} \leq 1$ & $n = 0,1,... < \left[\frac{1}{2\beta\lambda}\right]$ & $\epsilon_{n,0} = \sqrt{n} \Bigl[1-\beta \frac{\hbar e B_0}{2c} n \Bigr]^{\frac{1}{2}}$\\
\hline\hline
\end{tabular}
\label{table:V}
\end{table}

Using the energy spectrum and performing a comparison with the experimental results of the relativistic Landau levels in graphene \cite{zjiang}, we can estimate an upper bound on the deformed parameter $\beta_{\text{max}}$. The experiment uses an intense magnetic field of $B_0=18\ \text{Tesla}$ and Fermi velocity of $v_F = (1.12 \pm 0.02) \times 10^6 \text{m/s}$. The graphene Landau spectrum of the first excited level without the $\beta$-deformation is given by $E_{1}^{(\beta=0)} = v_F \sqrt{\frac{2\hbar e B_0}{c}} = (173 \pm 3) \text{meV}$ (experimental value). From \eqref{eigenE1}, we have
\begin{eqnarray}
E^{(\beta)}_{1} &=& E_{1}^{(\beta=0)}\sqrt{1-\beta \frac{\hbar e B_0}{2c}}.\label{eigenE3}
\end{eqnarray}
In the current experimental, it is impossible to distinguish the deviation induced by deformed quantum mechanics and thus 
\begin{eqnarray}
\left|\Delta E_{1}\right| &=& \left|E_{1}^{(\beta)} - E_{1}^{(\beta=0)}\right| = E_{1}^{(\beta=0)}\left|\left(\sqrt{1-\beta \frac{\hbar e B_0}{2c}}-1 \right)\right|\nonumber\\
 &=& E_{1}^{(\beta=0)} \left|\bigl(\sqrt{1-\delta} - 1\bigr)\right| < 6\ \text{meV}
\end{eqnarray} 
where we have defined $\delta=\beta \frac{\hbar e B_0}{2c}$. Since $\sqrt{1-\delta} < 1$, we obtain 
\begin{eqnarray}
1-\sqrt{1-\delta} < \frac{6}{173} \Rightarrow \delta < 0.068.
\end{eqnarray}
Therefore, the upper bound of the deformation parameter and thus the maximum momentum are deduced to be
\begin{eqnarray}
&& \beta_{\text{max}} < 4.5\times 10^{50}\ (\text{kg}^{-2} \text{m}^{-2} \text{s}^{2})\\
&& P_{\text{max}} =\frac{1}{\sqrt{\beta}} > 4.7\times 10^{-26} (\text{kg m} \text{s}^{-1}) \approx 87.9 \ \text{eV$/$c}. 
\end{eqnarray}
This bound is similar and consistent to the one derived in Snyder's algebra\cite{menculini-landau}. The constraints found above is not a very strong bound. However, it is consistent with\cite{das1}. In \cite{menculini-landau}, it was pointed out that the bound shall be improved and will become more stringent in future experiments where there is a possibility to measure Landau level transitions for very large quantum number $n$. Also, we point out that the deformed parameter $\beta$ may not be universal and could be system dependent\cite{brau}. In Sect.(4.1), we obtain a stronger bound by applying the modified dispersion relation induced from the anti-Snyder's model to the neutrino oscillation.

\section{Modified Energy Dispersion Relations (MEDR) and Time-dependent Speed of Light}

Recall from \eqref{MCR2}, we consider specific version of MCR $\left[X_{i}, P_{j}\right] = i\hbar\ \delta_{ij} (1-\beta P^2), \forall i,j=1,2,3$. In the exact representation in position space, we let $X^i = x^i$ and $P^i= f(p^i)$, we get
\begin{eqnarray}
\bigl\langle \bigl[X_i, f(p_j)\bigr]\bigr\rangle &=& i \hbar\ \delta_{ij} \bigl\langle \bigl(1-\beta f^2(p_k)\bigr)\bigr\rangle\nonumber\\
i\hbar \frac{\partial f(p_j)}{\partial p_i} &=& i \hbar\ \delta_{ij} \bigl(1-\beta f^2(p_k)\bigr)\Rightarrow \partial p_i = \frac{\partial f(p_i)}{1- \beta f^2(p_k)}\nonumber\\
\therefore P^i &:=& f(p^i) =\frac{1}{\sqrt{\beta}} \tanh\bigl(\sqrt{\beta} p^i\bigr) \label{exactP}\\
&=& p^{i}\ \bigl(1- \frac{\beta}{3} |p|^2 + O(\beta^2)\bigr).
\end{eqnarray}
Here $(x^{i},p^{i})$ are the usual low energy canonical pairs that satisfy the ordinary Heisenberg algebra, $\left[x^{i}, p^{j}\right]=i\hbar\ \delta^{ij}$ and $|p|=\sqrt{g_{ij} p^{i} p^{j}}$ is the magnitude of the three-momentum. 

Denoting $P^{\mu}$ as the four momentum\footnote{Greek indices run from $(\mu,\nu= 0,1,2,3)$ whereas italic indices run from $(i,j=1,2,3)$.}, we have
\begin{eqnarray}
X^{\mu} &=& (x^0, x^{i})\\
P^{\mu}&=&\Bigl(P^{0} = E/c,\ P^{i} =p^{i}\ \bigl(1- \frac{\beta}{3} |p|^2\bigr)\Bigr).\label{defP}
\end{eqnarray} 

Note that perturbatively we only keep terms up to $O(\beta)$. Next, we further denote the generic background metric as $g_{\mu\nu}$ (for simplicity assumed to be diagonal), the norm of $P^{\mu}$ is
\begin{eqnarray}
P^{\mu}P_{\mu} &=& g_{\mu\nu} P^{\mu} P^{\nu} = g_{00} \bigl(P^{0}\bigr)^2 + g_{ij} P^{i}P^{j}\nonumber\\
&=& g_{00}\frac{E^2}{c^2} + g_{ij} p^{i} p^{j}\ \bigl(1- \frac{\beta}{3} |p|^2\bigr)^2\nonumber\\
&=& g_{00}\frac{E^2}{c^2} + |p|^2 + |p|^2 \Bigl(-\frac{2\beta}{3} |p|^2 + O(\beta^2)\Bigr). 
\end{eqnarray}

Since the undeformed dispersion relation is $p^{\mu}p_{\mu}=g_{00} E^2/c^2 + |p|^2$ where $p^{\mu}p_{\mu}$ is the Casimir invariant of the usual Lorentz group given by the term $-m^2 c^2$, we obtain the modified dispersion relation in terms of low energy momentum
\begin{eqnarray}
P^{\mu}P_{\mu} &=& -m^2 c^2 -\frac{2\beta}{3} |p|^4. \label{disp}
\end{eqnarray}

One can invert \eqref{exactP} to express the undeformed momentum $p^{i}$ in terms of deformed $P^{i}$, i.e. $p^{i}= \frac{1}{\sqrt{\beta}} \tanh^{-1}\bigl(\sqrt{\beta}P^{i}\bigr) \approx P^{i}(1 + \frac{\beta}{3} |P|^2)$ where $|P|=\sqrt{g_{ij} P^{i}P^{j}}$. For simplicity, we keep to leading order $O(\beta)$. Thus, \eqref{disp} takes the form of 
\begin{eqnarray}
P^{\mu}P_{\mu} &=& -m^2 c^2 -\frac{2\beta}{3} |P|^4=- m_{\text{eff}}^2\ c^2\ ,\label{disp1}
\end{eqnarray}
where we have defined the modified effective mass as $m_{\text{eff}}:=\sqrt{m^2 + 2\beta |P|^4/(3c^2)}$. This correction can be viewed as the correction due to anti-Snyder physics that encoded maximum momentum scale in the model. Making used of $P^{\mu}P_{\mu}=g_{00} (P^{0})^2 + |P|^2$, Eq. \eqref{disp1} can be written as
\begin{eqnarray}
(P^{0})^2 &=& -\frac{1}{g_{00}}\left[m^2 c^2 + |P|^2 \Bigl(1 + \frac{2\beta}{3} |P|^2\Bigl)\right]\ ,\label{disp2}
\end{eqnarray}
and the energy of a particle in the gravitational field $g_{\mu\nu}$ \cite{padm} with anti-Snyder physics is
\begin{eqnarray}
E^2 &=& (-g_{00} c P^{0})^2= - g_{00} c^2 \left[m^2 c^2 + |P|^2 \Bigl(1 + \frac{2\beta}{3} |P|^2\Bigl)\right]\label{disp3}\\
&=& E_{0}^2 - \frac{2}{3} g_{00} c^2 \beta |P|^4.\nonumber
\end{eqnarray}    
Here $E_{0}^2=- g_{00} c^2 \bigl(m^2 c^2 + |P|^2\bigr)$ is the usual relativistic energy dispersion relation. If we consider Minkowski spacetime where $g_{00}=-1$ and set $m=0$, the photon's group velocity is defined by $v_{g}=\partial E/\partial |P|$. From \eqref{disp3} one can solve $|P|$ in terms of energy $E$ by iteration,     
\begin{eqnarray}
E^2 &=& c^2 |P|^2 \Bigl(1 + \frac{2\beta}{3} |P|^2\Bigr)\nonumber\\
\Rightarrow |P| &=& \pm \frac{E}{c}\left(1 - \frac{2\beta}{3}\frac{E^2}{c^2}\right).
\end{eqnarray}    
Hence up to $O(\beta)$, the modified photon group velocity is , 
\begin{eqnarray}
v_{g}&=&\partial E/\partial |P| = \pm c \left(1 + \frac{2\beta E^2}{c^2}\right)
\end{eqnarray} 
where $v_{g}= \pm c$ as $\beta\rightarrow 0$. Thus we have energy-dependent velocity of light at the Planck length regime with the new energy dispersion relation. This superluminal photon propagation, i.e. $v_{g}> c$ should be interpreted as the time-dependent velocity of light which is similar to the Varying Speed of Light (VSL) phenomena \cite{petit, moffat, magueijo} and also agree with DSR model \cite{amelino}. This superluminal phenomena appears in most of the UV completion of general relativity, for instance in massive gravity theory \cite{rham}. 

\subsection{Application to Neutrino Oscillations}

From \eqref{disp3} in Sect.(4), we discussed the deformed energy of a particle in the Minkowski spacetime with anti-Snyder algebra is given by
\begin{eqnarray}
E^2 &=& E_{0}^2 + \frac{2\beta}{3} c^2 |P|^4 \label{disp4}
\end{eqnarray}    
where $E_{0}^2= m^2 c^4 + c^2 |P|^2$ is the usual relativistic energy dispersion relation. Now, we apply this MEDR to study neutrino physics to obtain a bound on the deformation parameter $\beta$ through phenomenological study \cite{sprenger2}, say the recent experiments at MINOS experiment (Fermilab); Super-Kamiokande detection or Ice-Cube Neutrino Observatory. Neutrino oscillation refers to the modification of the Standard Model of particle physics to incorporate the neutrino flavour change/mixing. This is crucial to explain the solar neutrino problem \cite{gonzalez}. This novel phenomena accounts for the deficit in the number of electron neutrinos measured on Earth when compared to the prediction from theoretical models of the solar interior.

Neutrino oscillation between the flavours is possible through mixing and that neutrinos have mass. In other words, neutrino does not propagate in flavour eigenstates but in mass eigenstates. The basis change from the flavour eigenbasis to the mass eigenbasis is described by the Pontecorvo-Maki-Nakagawa-Sakata (PMNS) unitary matrix. This matrix is parametrized by three mixing angles and a Majorana CP-violating phase. The change of basis can be written as
\begin{eqnarray}
|\nu_{\alpha}\rangle &=&\sum_{i=1}^{3}U^{*}_{\alpha, i} |\nu_{i}\rangle
\end{eqnarray}      
where $|\nu_{\alpha}\rangle$ and $|\nu_{i}\rangle$ refer to flavour and mass eigenstates respectively. Since the free Hamiltonian operator is diagonal in mass eigenbasis, the mass eigenstates evolve through time and pick up an overall phase 
\begin{eqnarray}
|\nu_{i}(t)\rangle &=& e^{-iE_{i}t/\hbar}|\nu_{i}\rangle
\end{eqnarray}   
with $E_{i}$ the energy of the neutrino and $t$ the time of flight. The oscillation probability for a flavour change from flavour $\alpha$ to flavour $\beta$ is given by,
\begin{eqnarray}
P(\nu_{\alpha}\rightarrow \nu_{\beta})&=& |\langle \nu_{\beta}(t)|\nu_{\alpha}(t)\rangle|^2\nonumber\\
&=&\sum_{i,j=1}^{3} U^*_{\alpha,i} U_{\beta,i} U^*_{\alpha,j} U_{\beta,j}\ e^{-i c^4 \Delta m^2_{i,j} t/(2\hbar E)}
\end{eqnarray}
where $\Delta m^2_{i,j}\equiv m^2_{i} - m^2_{j}$ is the difference in the squared masses and $E$ the total energy of the system.

Let's consider the two-flavour oscillation in which the PMNS matrix is characterized by single mixing angle $\theta$,
\begin{eqnarray}
U &=& \begin{pmatrix} \cos\theta & \sin\theta \\ -\sin\theta & \cos\theta \end{pmatrix}
\end{eqnarray}
and the transition probability is reduced to
\begin{eqnarray}
P(\nu_{\alpha}\rightarrow \nu_{\beta})&=& \sin^2(2\theta) \sin^2\Bigl(\frac{(\Delta m^2)c^3 L}{4\hbar E}\Bigr) \label{neutp1}
\end{eqnarray}
where we assume that $E\approx c|\vec{P}_{i}|\approx c|\vec{P}_{j}|$ and denote $\Delta m^2=m_i^2-m_j^2$. The propagation length (baseline) is $L=c t$. If there is no mixing $\theta=0$, the probability above will be zero. Likewise, if the neutrinos are massless, then $\Delta m^2=0$, neutrino flavour change is also not possible. Experimentally, the difference in the squared masses of the neutrinos can be determined. One can further define the oscillation phase $\phi/(2\pi)=(\Delta m^2)c^3 L/(2\hbar E)=L/L_{0}$, such that the oscillation length $L_{0}=2\hbar E/((\Delta m^2)c^3)$. The transition probability takes the form
\begin{eqnarray}
P(\nu_{\alpha}\rightarrow \nu_{\beta})&=&\sin^2(2\theta) \sin^2\Bigl(\frac{\phi}{4\pi}\Bigr) =\sin^2(2\theta) \sin^2\Bigl(\frac{L}{2L_{0}}\Bigr).\label{neutp2}
\end{eqnarray}

Applying the MEDR, from \eqref{disp4} we have
\begin{eqnarray}
E &=& \sqrt{m^2 c^4 + c^2 |P|^2+ \frac{2\beta}{3} c^2 |P|^4} = c|P| \sqrt{1 + \frac{m^2 c^2}{|P|^2} + \frac{2\beta}{3} |P|^2}\nonumber\\
&\approx& c|P|\left(1 + \frac{m^2 c^2}{2|P|^2} + \frac{\beta}{3} |P|^2\right)
\end{eqnarray}
where we assume $m^2c^2 << |P|^2$ and keep to first leading order in $\beta$ during the expansion. Assuming $E\approx c|\vec{P}_{i}|\approx c|\vec{P}_{j}|$, the difference in energy for the mass eigenstates can be obtained as
\begin{eqnarray}
E_{i}-E_{j} &=& \frac{(\Delta m^2) c^4}{2E} + \frac{(\Delta \beta) E^3}{3c^2}.  
\end{eqnarray}
where $\Delta \beta= \beta_i - \beta_j$. The modified transition probability reads
\begin{eqnarray}
&&P^{\text{Anti-Snyer}}(\nu_{\alpha}\rightarrow \nu_{\beta}) = \sin^2(2\theta) \sin^2\left[\frac{(\Delta m^2)c^3 L}{4\hbar E} + \frac{(\Delta \beta) E^3 L}{6\hbar c^3}\right].\label{neutp3}
\end{eqnarray}
 
In \eqref{neutp3} if the deformed parameter $(\beta_{i}=\beta_{j})$ is the same for both eigenstates, then we need the standard non vanishing of difference of neutrino masses $(\Delta m^2)\neq 0$ to explain the neutrino oscillation phenomena. For $\beta_i \neq \beta_j$, comparing with the conventional neutrino oscillation in \eqref{neutp1}, the new oscillation phase is deviated by a factor that depends on the propagation length, the energy and the difference in the deformed parameter. This shift in the oscillating phase depends on $\text{sgn}(\Delta\beta)$. This effect is in contrast to the study of neutrinos oscillations with minimal length or in non-commutative geometry where in\cite{sprenger2} the authors found an exponential suppression of the oscillation phase that depends on energy, but not on the propagation length, given by
\begin{eqnarray}
&&P^{\text{Snyder}}(\nu_{\alpha}\rightarrow \nu_{\beta}) = \sin^2(2\theta) \sin^2\left[\frac{(\Delta m^2)c^3 L}{4\hbar E} \exp\Bigl(-\frac{(\Delta X)^2_{\text{min}}E^2}{c^2\hbar^2}\Bigr)\right]\nonumber\\
&& = \sin^2 (2\theta) \sin^2\left[1.27 (\Delta m^2[\text{eV}^2])\frac{L(\text{km})}{E([\text{GeV}])}\exp\Bigl(-\beta E^2[\text{GeV}^2]\Bigr)\right].
\end{eqnarray}
Here, $\beta$ is the deformed parameter in the unit of $\frac{c^2}{\text{GeV}^2}$. For comparison, in the Fig.(2), we plot the relative difference between classical oscillations, Snyder (minimal length) and anti-Snyder (maximal momentum, e.g. DSR-like theory) oscillations for the atmospheric neutrinos using the data from particle data group \cite{sprenger2, nakamura}. In this simplified case, we can model it as two-flavour oscillation in between neutrinos $\nu_{\mu}$ and $\nu_{\tau}$. We plot $P(\nu_{\mu} \rightarrow \nu_{\mu}) = 1 - P(\nu_{\mu} \rightarrow \nu_{\tau})$.
\begin{table}[ht]
\caption{Oscillation parameters used in the analysis of the modified oscillation behaviour in Sect.(4.1) and figure(3). We assume the CP-violting phase, $\delta_{\text{CP}}=0$.}
\centering
\begin{tabular}{c c c c c}
\hline\hline
Parameter & Value & Parameter & Value & Experiment\\ [1ex] \hline
$\sin^2(2\theta_{1,2})$ & 0.861 , & $\Delta m^2_{1,2}$ & $7.59\times 10^{-5}\text{eV}^2$ & Solar\\ [1ex] \hline
$\sin^2(2\theta_{1,3})$ & 0.15 , & $\Delta m^2_{1,3}$ & $2.43\times 10^{-3}\text{eV}^2$ & Cross-mixing\\[1ex] \hline
$\sin^2(2\theta_{2,3})$ & 0.92 , & $\Delta m^2_{2,3}$ & $2.43\times 10^{-3}\text{eV}^2$ & Atmospheric\\ [1ex]\hline\hline
\end{tabular}
\label{table:VI}
\end{table}

In Fig.(2) we see that the minimal length suppresses the oscillations exponentially for all finite values of $\Delta X_{\text{min}}=\hbar\sqrt{\beta}$. In contrast, the anti-Snyder case is drastically different compared to the Snyder (minimal length) case because the modification is caused by $\Delta\beta$ instead of $\beta$. We expect the smallness of the non-universality of $\beta$, i.e. $\Delta \beta<< 1$. This is consistent with the Fig. (2) since we observe that when $\Delta\beta$ is increasing, the extremely rapid oscillation will wash out the neutrino oscillation. Thus, neutrino oscillations can provide novel information on Snyder or anti-Snyder induced phenomena.

From another perspective, it is possible for neutrino oscillation with massless or degenerate mass neutrinos if $\Delta \beta= \beta_i - \beta_j \neq 0$, i.e. deformed parameters are flavour dependent. In this case, we compare the conventional theory with the deformed case, we get
\begin{eqnarray}
\frac{(\Delta m^2)c^3 L}{4\hbar E} &=& \frac{(\Delta \beta) E^3 L}{6\hbar c^3}\nonumber\\
\Rightarrow (\Delta \beta) &=& \frac{3(\Delta m^2)c^6}{2E^4} \approx 5.25 \times 10^{30} \frac{\bigl(\Delta m^2[\text{eV}^2]\bigr)}{E^4[\text{MeV}^4]}.
\end{eqnarray}
From the (solar) neutrino oscillation in vacuum, the fit for the difference in mass-squared is $(\Delta m_{1,2}^2) \approx 7.59\times 10^{-5} \text{eV}^2$ \cite{boris}. The energy ranges from $E \approx 0.4 \rightarrow 18\ \text{MeV}$ depending on the mechanisms. Thus, we have largest $\Delta \beta \approx 1.57 \times 10^{28}$ in S.I unit. Thus, from solar neutrino physics, we have the constraint(upper bound) on the deformed parameter as 
\begin{eqnarray}
\beta^{\text{sol}}_{\text{max}} \leq 1.57 \times 10^{28}\ (\text{kg}^{-2} \text{m}^{-2} \text{s}^2). 
\end{eqnarray}
Likewise, for the atmospheric neutrino data, we have $(\Delta m_{23}^2) \approx 2.43\times 10^{-3} \text{eV}^2$ and $E \approx 1\ \text{GeV}$, the bound is
\begin{eqnarray}
\beta^{\text{atm}}_{\text{max}} \leq 1.27 \times 10^{16}\ (\text{kg}^{-2} \text{m}^{-2} \text{s}^2). 
\end{eqnarray}

Here we have more stringent bounds as compared to the one by Landau levels in graphene in Sect.(3.2). In the literatures, there are few bounds derived from low energy physics for the quadratic GUP (Snyder model) \cite{das1}: $\beta < 10^{50}$ (Landau levels); $\beta < 10^{36}$ (Lamb Shift), $\beta < 10^{21}$ (STM current). Meanwhile for linear GUP \cite{das} we have $\alpha < 10^{23}$ (Landau levels); $\alpha < 10^{17}$ (SHO), $\alpha < 10^{10}$ (Lamb Shift). On dimensional ground, $[\beta]=[\alpha]^2$. We see that our constraints are consistent with the Snyder case. The well-known electroweak energy scale is $E_{\text{ew}} \approx 0.25\ \text{TeV}$. In our case, the lower bound of the maximum momentum/energy corresponding to the largest $\beta$ is about $E_{\text{cutoff}} \geq \frac{c}{\sqrt{\beta}} \approx 14.6\ \text{TeV} > E_{\text{ew}}$. Therefore, it could signal a new and intermediate energy scale between the electroweak and the Planck scale which makes the quantum gravity phenomenology interesting enough to be further explored.

\section{Conclusion}

We have studied the non-relativistic anti-Snyder models which can be viewed as realization of DSR axioms and it's possible applications in low energy physics. After formulating the Hilbert space in momentum representation, we have discussed the symmetricity condition of the physical operators. In the deformed space, both deformed position and momentum operator are essentially self-adjoint by Von-Neumann's theorem.

After casting the (2+1)-dimensional Dirac equation with homogeneous magnetic fields in terms of Schr$\ddot{\text{o}}$dinger-like equation with complexified Hyperbolic Scarf potential, we have obtained exact solution of the relativistic (2+1) dimensional Dirac-Landau problem. The exact solutions are given by orthogonal Romanovski polynomials and only finite number of possible bound state solutions are permitted. The eigenenergy spectrum is modified with negative shift as compared to the standard relativistic Dirac-Landau problem. These are the crucial effects due to the maximum momentum/energy that are introduced by DSR-like theory. We classify the physical solutions accordingly to the angular momentum quantum number $m$. Both table 3 $\bigl(\frac{1}{\beta\lambda} > 1\bigr)$ and table 4 $\bigl(\frac{1}{\beta\lambda}\leq 1\bigr)$ give the states which are valid for all angular quantum number with infinite degeneracy. Thus they can be considered as solutions closest to the standard undeformed Dirac-Landau problem with the deformed energies. Besides that, we have also obtained another two classes of solutions that depend on angular momentum, i.e. Table 1 $\bigl(-\frac{1}{2}+\frac{1}{2\beta\lambda} < m \leq \frac{1}{\beta\lambda}\bigr)$ and Table 2 $\bigl(-1 \leq m < -\frac{1}{2}+\frac{1}{2\beta\lambda}\bigr)$ with finite degeneracy. Angular momentum quantum of these states are constraint within the range $-1\leq m\leq \frac{1}{\beta\lambda}$. In the limit $\beta\rightarrow 0$, the upper bound of the constraint on quantum number $m$ disappears and table 1 becomes meaningless and is no longer physically acceptable. However, table 2 becomes valid for $-1 \leq m < \infty$. Thus, in the limit of vanishing of maximum momentum cutoff $P_{\text{max}}\rightarrow \infty$, both table 2 and table 3 reproduce the correct physical states of ordinary quantum mechanics.

We have discussed the massless charged fermion case and compared the energy correction with the experimental results on lowest Landau levels. We obtained an upper bound on deformed parameter similar to the Snyder case, $\beta_{\text{max}} < 4.5\times 10^{50}\ (\text{kg}^{-2} \text{m}^{-2} \text{s}^{2})$. Heuristically, it means there is a lower bound on the maximum momentum cutoff $P_{\text{max}} = \frac{1}{\sqrt{\beta}} > \approx 87.9\ \text{eV$/$c}$.

Next, we have examined the modified energy dispersion relation induced by the anti-Snyder model. The group velocity of photon is found to be time-dependent and superluminal. We interpret this as varying speed of light phenomena. Subsequently, by applying the modified dispersion relation to neutrinos oscillation, we have obtained the modified oscillating phase and compared it to the Snyder model. In Snyder model, the oscillating phase is exponentially suppressed. However, in anti-Snyder model the modification is only possible if the deformed parameter is not universal and is flavour dependent. Interestingly, this non-universality allows us to consider the neutrino oscillation for massless and mass-degenerate neutrinos. Finally, by comparing with Particle data, we have obtained a more stringent bound on the deformed parameter, $\beta^{\text{sol}}_{\text{max}} \leq 1.57 \times 10^{28}\ (\text{kg}^{-2} \text{m}^{-2} \text{s}^2)$ for solar neutrinos and $\beta^{\text{atm}}_{\text{max}} \leq 1.27 \times 10^{16}\ (\text{kg}^{-2} \text{m}^{-2} \text{s}^2)$ for atmospheric neutrinos.

\section{Acknowledgment}
C.L. Ching thanks Rajesh Parwani and Kuldip Singh for encouragement.

\section{Appendix: Hypergeometric Equation and Finite Romanovski Polynomial}

All classical orthogonal polynomials appear as solutions of hypergeometric equation as shown below:
\begin{eqnarray}
\sigma(x) \psi''_{n} (x) + \tau(x) \psi'_{n} (x) -\lambda_{n} \psi_{n} (x)=0
\end{eqnarray}
where the coefficients are
\begin{eqnarray}
\sigma(x) &=& ax^2 + bx +c\nonumber\\
\tau(x) &=& dx + e \nonumber\\
\lambda_n &=& n(n-1)a + nd. \nonumber
\end{eqnarray}
There are various methods to find the solution \cite{alvarez1, alvarez2} and typically denoted by
\begin{eqnarray}
\psi_{n} (x)= P_n \left(\begin{matrix}
 d &  & e \\
 a & b & c 
\end{matrix}\middle|x\right)
\end{eqnarray}
in which $(a,b,c,d,e)$ have been made explicit as the equation parameters of the polynomial of degree $n(=0,1,2,...)$, and $\lambda_n$ is the eigenvalue parameter. From Nikiforov and Uvarov method \cite{nikiforov}, the polynomials $\psi_n (x)$ can be built from Rodrigues representation and are classified according to their weight functions $w(x)$ as following:
\begin{eqnarray}
P_n(x) &=& \frac{1}{W\left(\begin{matrix}
 d &  & e \\
 a & b & c 
\end{matrix}\middle|x\right)} \times \frac{d^n}{dx^n}\left[(ax^2 + bx +c)^n W\left(\begin{matrix}
 d &  & e \\
 a & b & c 
\end{matrix}\middle|x\right)\right]\\
w(x) &=& \mathcal{W}\left(\begin{matrix}
 d &  & e \\
 a & b & c 
\end{matrix}\middle|x\right) = \exp\left(\int dx \frac{(d-2a)x + (e-b)}{ax^2 + bx +c}\right).
\end{eqnarray}
By choosing appropriate parameters, one can obtain orthogonal solutions: Jacobi (include special cases likes Legendre, Gegenbauer and Chebychev), Laguerre, Hermite polynomials and non-orthogonal solution: Bessel polynomials which are familiar in mathematical physics literature. However, there is another class of solution which is less well known as ``finite Romanovski polynomials" denoted by $\psi_{n}(x) = R^{(p,q)}_n(x)$. The corresponding set of parameters are $a=1, b=0, c=1, d=2(1-p)$ and $e=q$ where $p>0$. Thus, the differential equation satisfied by the Romanovski polynomials is given by
\begin{eqnarray}
\left[(1 + x^2)\frac{d^2}{dx^2} + (2 (-p+1) x + q)\frac{d}{dx} - (n (n-1) +2n (1-p))\right] R^{(p,q)}_n(x) = 0.
\end{eqnarray}
Such a set of parameters results in the following weight function $w^{(p,q)}(x)$ that belongs to this class of polynomial,
\begin{eqnarray}
w^{(p,q)}(x) &=& (1 + x^2)^{-p} e^{q\tan^{-1} x}
\end{eqnarray}
and with non-trivial orthogonality condition
\begin{eqnarray}
\int_{-\infty}^{+\infty} dx\ w^{(p,q)}(x)\ R_n^{(p,q)}(x)\ R_{n'}^{(p,q)}(x) < \infty
\end{eqnarray}
if and only if $n+n' < 2p-1$.

\section*{Figure Captions}

\begin{itemize}

\item  Figure 1: The deformed Landau levels $\epsilon (n,\beta) = \frac{E_{n}}{v_F\sqrt{2\hbar e \ B_0/c}}$ for both Snyder (purple) /anti-Snyder (red) models. Here we set $\lambda=\frac{\hbar e B_0}{2c}=1; \beta=0.01$. Dashed line is the landau levels for the standard quantum mechanics.

\item  Figure 2: The classical (dashed line), Snyder with minimal length (purple) and anti-Snyder with maximum momentum (red) deformed oscillations for the atmospheric neutrinos. It can be modeled as two-flavour case in between neutrinos $\nu_{\mu}$ and $\nu_{\tau}$. We set $(\Delta m^2_{2,3})= 2.43\times 10^{-3}\ \text{eV}^2$, $\sin^2(2 \theta_{23}) = 0.92$, $L=5000\ \text{km}$, $\beta =0.01 c^2/(\text{GeV}^2)$ and $\Delta\beta \approx 10^{-27} c^2/(\text{GeV}^2)$.

\end{itemize}

\section*{Figures}

\begin{figure}[ht]
  \begin{center}
   \epsfig{file=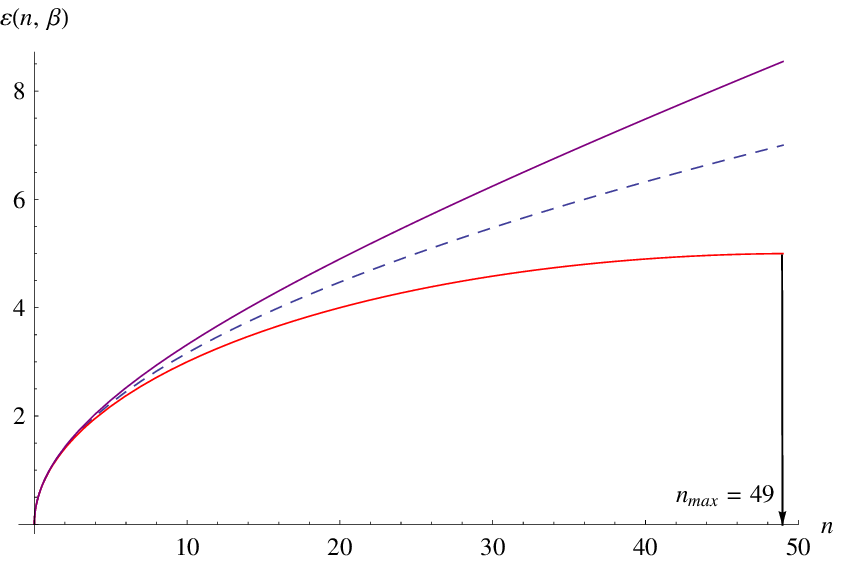, width=12cm}
    \caption{}
    \label{fig1}
  \end{center}
\end{figure}

\begin{figure}[ht]
  \begin{center}
   \epsfig{file=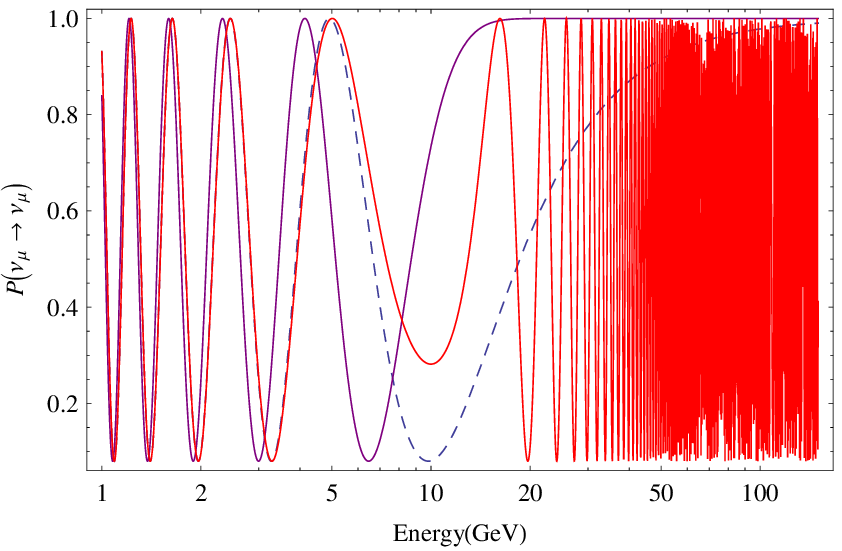, width=12cm}
    \caption{}
    \label{fig2}
  \end{center}
\end{figure}

\end{document}